\documentclass [12pt]{article}
\usepackage {graphicx}

\begin{document}

\title{Addressing the bias in Monte Carlo pricing of multi-asset options with
multiple barriers through discrete sampling}

\author{Pavel V. Shevchenko\footnote{ contact e-mail: Pavel.Shevchenko@csiro.au}\\
CSIRO Mathematical and Information Sciences, Sydney, Australia }

\date{27 March 2002}

\maketitle

\begin{center}
This is a preprint of an article published in \\The
Journal of Computational Finance 6(3), pp.1-20, 2003.\\
www.journalofcomputationalfinance.com
\end{center}

\begin{abstract}
An efficient conditioning technique, the so-called Brownian Bridge
simulation, has previously been applied to eliminate pricing bias that
arises in applications of the standard discrete-time Monte Carlo method to
evaluate options written on the continuous-time extrema of an underlying
asset. It is based on the simple and easy to implement analytic formulas for
the distribution of one-dimensional Brownian Bridge extremes. This paper
extends the technique to the valuation of multi-asset options with knock-out
barriers imposed for all or some of the underlying assets. We derive formula
for the unbiased option price estimator based on the joint distribution of
the multi-dimensional Brownian Bridge dependent extrema. As analytic
formulas are not available for the joint distribution in general, we develop
upper and lower biased option price estimators based on the distribution of
independent extrema and the Fr\'{e}chet lower and upper bounds for the
unknown distribution. All estimators are simple and easy to implement. They
can always be used to bind the true value by a confidence interval.
Numerical tests indicate that our biased estimators converge rapidly to the
true option value as the number of time steps for the asset path simulation
increases in comparison to the estimator based on the standard discrete-time
method. The convergence rate depends on the correlation and barrier
structures of the underlying assets.
\end{abstract}

\noindent
Key Words: Monte Carlo simulation, extreme values,
Brownian Bridge, multi-asset barrier option, multi-variate joint
distribution, Fr\'{e}chet bounds.

\section{Introduction}

Barrier options introduced by Merton (1973) are used widely in trading now.
The option is extinguished (knocked-out) or activated (knocked-in) when an
underlying asset reaches a specified level (barrier). A lot of related more
complex instruments for example bivariate barrier, ladder, step-up or
step-down barrier options have become very popular in over-the-counter
markets. In general, these options can be considered as options with payoff
depending upon the path extrema of the underlying assets. A variety of
closed form solutions for such instruments on a single underlying asset have
been obtained in the classical Black-Scholes settings of constant
volatility, interest rate and barrier level. See for example Heynen and Kat
(1994a), Kunitomo and Ikeda (1992), Rubinstein and Reiner (1991). If the
barrier option is based on two assets then a practical analytical solution
can be obtained for some special cases considered in Heynen and Kat (1994b)
and He, Keirstead and Rebholz (1998).

In practice, however, numerical methods are used to price the barrier
options for a number of reasons, for example, if the assumptions of constant
volatility and drift are relaxed or payoff is too complicated. Numerical
schemes such as binomial and trinomial lattices (Hull and White (1993), Kat
and Verdonk (1995)) or finite difference schemes (Dewynne and Wilmott
(1993)) can be applied to the problem. However, the implementation of these
methods can be difficult. Also, if more than two underlying assets are
involved in the pricing equation then these methods are not practical.

In this paper we focus on a Monte Carlo simulation method which is
a good general pricing tool for such instruments. However, finding
the extrema of the continuously monitored assets by sampling
assets at discrete dates, the standard discrete-time Monte Carlo
approach, is computationally expensive as a large number of
sampling dates and simulations are required. Loss of information
about all parts of the continuous-time path between sampling dates
introduces a substantial bias for the option price. The bias
decreases very slowly as $1 / \sqrt M $ for $M>>1$, where $M$ is
the number of equally spaced sampling dates (see Broadie,
Glasserman and Kou (1997)). Also, extrapolation of the Monte Carlo
estimates to the continuous limit is usually difficult due to
finite sampling errors. For the case of a single underlying asset,
it was shown by Andersen and Brotherton-Ratcliffe (1996) and
Beaglehole, Dybvig and Zhou (1997) that the bias can be eliminated
by a simple conditioning technique, the so-called Brownian Bridge
simulation. The method is based on the simulation of a
one-dimensional Brownian bridge extremum between the sampled dates
according to a simple analytical formula for the distribution of
the extremum. The technique is very efficient because only one
time step is required to simulate the asset path and its extremum
if the barrier, drift and volatility are constant over the time
region. We extend the technique to the valuation of multi-asset
options with continuously monitored knock-out barriers imposed for
some or all underlying assets. We derive the general formula for
the unbiased estimator based on the joint distribution of the
multi-dimensional Brownian Bridge dependent extrema. In general,
however, the analytic formulas are not available for this joint
distribution and we develop three biased estimators. The upper and
lower estimators are based on the Fr\'{e}chet bounds for the
unknown multi-variate joint distribution of the extrema. The third
estimator (which is typically most accurate) is based on the joint
distribution of the independent extrema. The biased estimators can
be used to bind the true option price by a confidence interval.
Numerical examples indicate that the biases rapidly decrease as
$M$ increases in comparison to the bias in the standard
discrete-time method and the convergence rate depends on the
correlation and barrier structures of the underlying assets.
Finally we discuss the application of our biased estimators for
the valuation of knock-in, cash-at-hit rebates options, lookback
options and credit derivatives.

\section{Unbiased estimator via Brownian Bridge correction}

\subsection{Model setup}

Consider the knock-out option $Q$ written on the underlying assets
$\vec {S}(t) = (S_1 (t),...,S_d (t))$. The option payout at
maturity $t = T$ is some function $V(\vec {S}(T))$  if the
underlying assets never hit the fixed boundaries $\vec {h}(t) <
\vec {S}(t) < \vec {H}(t),\;\;t \in [0,T]$ and zero otherwise.
Here $\vec {H}(t) = (H_1 (t),...,H_d (t))$ and $\vec {h}(t) = (h_1
(t),...,h_d (t))$ are the upper and lower barriers. Hereafter, we
use vector notation to simultaneously compare all vector
components. For example, $\vec {A} < \vec {B}$ is used to denote
$A_i < B_i $ for all $i = 1,...,d$.

Assume that the underlying assets follow risk-neutral geometric Brownian
motion

\begin{equation}
\label{eq1}
dS_i (t) / S_i (t) = \mu _i (t)dt + \sigma _i (t)dW_i (t),\quad E[dW_i
(t)dW_j (t)] = \rho _{ij} (t)dt,
\end{equation}

\noindent
where $S_i (0)$ is the $i$-th asset price today, $W_i (t),i = 1,...,d$ is the
$d$-dimensional Wiener process, $\vec {\mu }(t) = (\mu _1 (t),...,\mu _d (t))$
and $\vec {\sigma }(t) = (\sigma _1 (t),...,\sigma _d (t))$ are the drifts
and volatilities respectively. Let us consider the time slices $t_m $, $\;m
= 0,...,M$ ordered as $0 = t_0 < t_1 < t_2 < ... < t_M = T$, $\delta t_m =
t_{m + 1} - t_m $ and denote $S_i (t_m ) = S_i^{(m)} $. We assume that the
drifts, volatilities, correlation coefficients and boundaries are piecewise
constant functions of time such that $\mu _i (t) = \mu _i^{(m)} $, $\sigma
_i (t) = \sigma _i^{(m)} $, $\rho _{ij} (t) = \rho _{ij}^{(m)} $, $H_i (t) =
H_i^{(m)} $, $h_i (t) = h_i^{(m)} $ if $t \in [t_m ,t_{m + 1} )$, $m =
0,1,...,M - 1$. Let us also introduce the indicator function of the barrier
hit at discrete times $t_m ,\;m = 0,...,M$ by

\begin{equation}
\label{eq2}
{\rm I}_{\tau > T} = \left\{ {{\begin{array}{*{20}c}
 {1,\quad \mbox{if}\quad \vec {h}^{(m)} < \vec {S}^{(m)} < \vec
{H}^{(m)}\;\mbox{for}\;m = 0,...,M,} \\
 {0,\quad \mbox{otherwise.}} \\
\end{array} }} \right.
\end{equation}

In the absence of arbitrage the true option price at $t = 0$ can be written
as an expectation

\begin{equation}
\label{eq3}
Q = \int\limits_{\vec {h}^{(\ref{eq1})}}^{\vec {H}^{(\ref{eq1})}} {d\vec {S}^{(\ref{eq1})}}
...\int\limits_{\vec {h}^{(M)}}^{\vec {H}^{(M)}} {d\vec {S}^{(M)}V(\vec
{S}^{(M)})p(\vec {S}^{(\ref{eq1})}\vert \vec {S}^{(0)})} \cdot ... \cdot p(\vec
{S}^{(M)}\vert \vec {S}^{(M - 1)}),
\end{equation}

\noindent
where $p(\vec {S}^{(m + 1)}\vert \vec {S}^{(m)})$ is the \textit{risk-neutral} probability
density function of the asset value $\vec {S}^{(m + 1)}$ at $t_{m + 1} $
given the asset value $\vec {S}^{(m)}$ at $t_m $. The function should
satisfy the Kolmogorov forward equation (also known as the Fokker-Planck
equation) with the absorbing boundaries $\vec {h}(t),\;\vec {H}(t),\;\;t \in
[0,T]$, see Cox and Miller (1965). Also we have absorbed the present value
discount factor into the payoff function $V(\vec {S}(T))$ and used the short
vector notation for the multi-dimensional integral\footnote{
$\int\limits_{\vec {h}^{(m)}}^{\vec {H}^{(m)}} {d\vec {S}^{(m)}} =
\int\limits_{h_1^{(m)} }^{H_1^{(m)} } {dS_1^{(m)} ...}
\int\limits_{h_d^{(m)} }^{H_d^{(m)} } {dS_d^{(m)} } $}.

The explicit solution of the Kolmogorov forward equation for the transition
probability function $p_0 (\vec {S}^{(m + 1)}\vert \vec {S}^{(m)})$ to
unrestricted process (\ref{eq1}) over the interval $[t_m ,t_{m + 1} ]$ (without
absorbing boundaries) is the $d$-variate lognormal distribution with the means:
$[\mu _i^{(m)} - 0.5(\sigma _i^{(m)} )^2]\delta t_m - \ln S_i ^{(m)}$,
variances: $(\sigma _i^{(m)} )^2\delta t_m $, and linear correlation
coefficients: $\rho _{ij}^{(m)} $, $i = 1,...,d;\;j = 1,...,d$. The
simulation of $S_i^{(m + 1)} $ from this distribution is simply

\begin{equation}
\label{eq4}
S_i^{(m + 1)} = S_i^{(m)} \exp \left\{ {[\mu _i^{(m)} - 0.5(\sigma _i^{(m)}
)^2]\delta t_m - \sigma _i^{(m)} \sqrt {\delta t_m } Z_i^{(m)} } \right\},
\end{equation}

\noindent
where $Z_i^{(m)} ,\;i = 1,...,d$ are random variates from the $d$-variate
Normal distribution with the linear correlation coefficients $\rho
_{ij}^{(m)} $, zero means and unit variances for given $m$ (the random variates
are independent for different $m)$.

\subsection{Monte Carlo estimators}

The standard discrete-time Monte Carlo approach to estimate the knock-out
barrier options (\ref{eq3}) is to assume unrestricted process between sampling dates
$t_m ,\;\;m = 0,1,..,M$, then simulate $N$ independently drawn asset paths
according to the iterative equation (\ref{eq4}) and finally to calculate the option
price estimate as

\begin{equation}
\label{eq5}
Q_S = V(\vec {S}(T)){\rm I}_{\tau > T} .
\end{equation}

\noindent
For simplicity, hereafter, we omit the averaging over
$N$ paths. Finding the option price according to (\ref{eq5}) will
introduce a bias\footnote{ The estimator (\ref{eq5}) is an
unbiased estimator of the option with discretely monitored
barriers if sampling dates match the barrier monitoring dates.}
which is usually larger than the statistical error of the Monte
Carlo estimates. This is because we lose information on the
continuous-time path between sampling dates. The bias decreases
very slowly as $Q_S - Q\sim 1 / \sqrt M $ for $M >
> 1$ (see Andersen (1996) and Broadie, Glasserman and Kou (1997)). Thus a
large number of sampling dates is usually required to obtain an
accurate estimate of the option price with continuously monitored
barriers. For example, the bias is still larger than 1{\%} of the
true price even for 1024 time steps for the case of the standard
down-and-out call, see Table 1. Extrapolation to the continuous
limit is complicated by finite sampling errors of the Monte Carlo
estimates. This makes the estimator (\ref{eq5}) computationally
expensive. Being mainly interested in eliminating (reducing) the
biases unaffected by the number of paths, hereafter in formulae,
we omit the dependence of the Monte Carlo estimates upon $N$
assuming that $N$ is large enough to make the statistical errors
negligibly small in comparison to the biases (we will present the
standard errors in the numerical examples).

Consider the continuous-time asset maxima, $\vec {M}^{(m)} = (M_1^{(m)}
,...,M_d^{(m)} )$, and minima, $\vec {L}^{(m)} = (L_1^{(m)} ,...,L_d^{(m)}
)$, over the time interval $[t_m ,t_{m + 1} ]$, where

\begin{equation}
\label{eq6}
M_i^{(m)} = \max \{S_i (t):t \in [t_m ,t_{m + 1} ]\},\;\;L_i^{(m)} = \min
\{S_i (t):t \in [t_m ,t_{m + 1} ]\}.
\end{equation}

\noindent
If the continuous barriers are imposed between sampling
dates then the correct transition probability function (which is a
solution of the Kolmogorov forward equation with absorbing
boundaries)

\begin{equation}
\label{eq7}
p(\vec {S}^{(m + 1)},\vec {M}^{(m)} < \vec {H}^{(m)},\vec {L}^{(m)} > \vec
{h}^{(m)}\vert \vec {S}^{(m)}),
\end{equation}

\noindent
should be used in the option price integral (\ref{eq3}) rather than the transition
probability function $p_0 (\vec {S}^{(m + 1)}\vert \vec {S}^{(m)})$ for an
unrestricted process. Once the asset path is simulated\footnote{ The asset
path should not necessarily be simulated according to (\ref{eq4}). In general it can
be simulated from any distribution.} according to (\ref{eq4}) then the unbiased
estimator for the knock-out option can be calculated as

\begin{equation}
\label{eq8}
Q = V(S(T)) \cdot {\rm I}_{\tau > T} \prod\limits_{m = 1}^M {P^{(m)}} ,
\end{equation}

\noindent
where

\begin{eqnarray}
\label{eq9}
 P^{(m)} = \frac{p(\vec {S}^{(m + 1)},\;\vec {M}^{(m)} < \vec {H}^{(m)},\vec
{L}^{(m)} > \vec {h}^{(m)}\vert \vec {S}^{(m)})}{p_0 (\vec {S}^{(m
+ 1)}\vert \vec {S}^{(m)})}  \\ \nonumber
 \quad \quad \, = \mbox{Pr}[\vec {M}^{(m)} < \vec {H}^{(m)},\vec {L}^{(m)} >
\vec {h}^{(m)}\vert \vec {S}^{(m + 1)},\vec {S}^{(m)}]
\end{eqnarray}

\noindent is the probability that the assets will not hit the
barriers in the time region $[t_m ,t_{m + 1} ]$ conditional on
$\vec {S}^{(m)}$, $\vec {S}^{(m + 1)}$. This probability is a
joint distribution of the extrema $\vec {L}^{(m)}$ and $\vec
{M}^{(m)}$ conditional on $\vec {S}^{(m)}$, $\vec {S}^{(m + 1)}$.
After suitable normalization\footnote{ The log-increments of $\vec
{S}(t)$ conditional on $\vec {S}^{(m)}$ and $\vec {S}^{(m + 1)}$
become a so-called Brownian Bridge on $[t_m ,t_{m + 1} ]$.}
$P^{(m)}$ is a joint distribution of the Brownian Bridge extrema
(see Karatzas and Shreve (1991) or Borodin and Salminen (1996)).
If this probability is easy to calculate then the estimator
(\ref{eq8}) is trivial to implement. One has to simulate the asset
path at $t_m $, $m = 0,1,...,M$ according to the standard
procedure (\ref{eq4}). If the underlying assets never hit the
barriers at the sampling dates then the option price estimator is
given by the discounted payoff $V(\vec {S}(T))$ weighted with
$\prod\limits_{m = 1}^M {P^{(m)}} $ and zero otherwise.  However,
joint distribution $P^{(m)}$ can be found analytically for some
special cases only. In the case of barriers discretely monitored
at sampling dates, $P^{(m)} = 1$ and estimator (\ref{eq8}) is
equivalent to the discrete-time estimator (\ref{eq5}).

\subsection{Marginal distributions of the Brownian Bridge extrema}

For the case of a single barrier for one of the assets at each
time region the probability $P^{(m)}$ in (\ref{eq9}) can be found
analytically. Using results from the theory of the Wiener process
with absorbing boundaries, Cox and Miller (1965), and (\ref{eq9})
or the formula for the one-dimensional Brownian Bridge extremum,
Karatzas and Shreve (1991), the marginal distributions of the
maximum $M_k^{(m)} $ and minimum $L_k^{(m)} $ of the $k$-th asset
conditional on $\vec {S}^{(m)}$ and $\vec {S}^{(m + 1)}$ are given
by

\begin{equation}
\label{eq10}
\begin{array}{l}
 \Pr [M_k^{(m)} < H_k^{(m)} \vert \vec {S}^{(m)},\vec {S}^{(m + 1)}] = 1 -
\xi _k^{(m)} (H_k^{(m)} ),\; \\
 \;\Pr [L_k^{(m)} > h_k^{(m)} \vert \vec {S}^{(m)},\vec {S}^{(m + 1)}] = 1 -
\xi _k^{(m)} (h_k^{(m)} ), \\
 \end{array}
\end{equation}

\noindent
where

\begin{equation}
\label{eq11}
\begin{array}{l}
 h_k^{(m)} < \min [S_k^{(m + 1)} ,S_k^{(m)} ],\;H_k^{(m)} > \min [S_k^{(m +
1)} ,S_k^{(m)} ], \\
 \xi _k^{(m)} (X) = \exp \left( { - 2\ln \textstyle{X \over {S_k^{(m)} }}\ln
\textstyle{X \over {S_k^{(m + 1)} }} / (\sigma _k^{(m)} )^2\delta t_m }
\right)\;. \\
 \end{array}
\end{equation}

\noindent Here $\xi _k^{(m)} (h_k^{(m)} )$ and $\xi _k^{(m)}
(H_k^{(m)} )$ are the marginal probabilities of the upper and
lower barrier hits by the $k$-th asset in the interval $[t_m ,t_{m
+ 1} ]$ respectively. The maximum and minimum can be simulated
marginally by $M_k^{(m)} = (\xi _k^{(m)} )^{ - 1}(1 - U)$ and
$L_k^{(m)} = (\xi _k^{(m)} )^{ - 1}(U)$, where $U$ is a random
variable from the standard Uniform distribution\footnote{ The
inverse function $(\xi _k^{(m)} )^{ - 1}(U)$ has two solutions.
One solution is used to find the maximum and the other is used to
find the minimum.}. Thus, if the only barrier at $[t_m ,t_{m + 1}
]$ is imposed for the $k$-th asset then $P^{(m)} = 1 - \xi
_k^{(m)} (h_k^{(m)} )$ for the case of a lower barrier and
$P^{(m)} = 1 - \xi _k^{(m)} (H_k^{(m)} )$ for the case of an upper
barrier. The marginal distributions (\ref{eq10}) are valid as long
as the interest rate, asset volatility and drift are constant in
the time interval where the constant barrier is imposed on the
asset.

\subsection{Single barrier option}

For a single underlying asset and single barrier per time region it was
demonstrated by Andersen and Brotherton-Ratcliffe (1996) and Beaglehole,
Dybvig and Zhou (1997) that simulation of the barrier hits in the interval
$[t_m ,t_{m + 1} ]$ using (\ref{eq10}) eliminates the bias presented in (\ref{eq5}).
Alternatively we calculate the option price estimator using (\ref{eq8}). We would
like to stress that the marginal probabilities (\ref{eq10}) can be used to get the
unbiased option price estimator not only for single asset barrier option but
also for multi-asset options if there is a single barrier at each time
region (this barrier can be imposed for different assets at different time
regions).

In Table 1 we present the Monte Carlo results for down-and-out
call for the cases of one and two underlying assets. In the first
case the option pays $\max [S_1 (T) - K,\;0]$ if $S_1 (t) > h_1
,\;t \in [0,T]$ and in the second case the option pays $\max [S_1
(T) - K,\;0]$ if $S_2 (t) > h_2 ,\;t \in [0,T]$. We have
calculated the standard discrete-time biased estimator, $Q_S $,
using (\ref{eq5}) and the unbiased estimator, $Q$, using
(\ref{eq8}) versus the number of equally spaced time steps $M$.
All parameters: volatilities, drifts, barriers, correlations are
assumed constant. Explicit formulae for the unbiased estimators in
these examples are

$$
Q = e^{ - rT}\max [S_1 (T) - K,0] \cdot {\rm I}_{\tau > T}
\prod\limits_{m = 1}^M {[1 - \xi _1^{(m)} (h_1 )]}
$$

\noindent
for the case of a single underlying asset and

$$
Q = e^{ - rT}\max [S_1 (T) -
K,0] \cdot {\rm I}_{\tau > T} \prod\limits_{m = 1}^M {[1 - \xi
_2^{(m)} (h_2 )]}
$$

\noindent
for the case of two underlying assets. Being mainly
interested in eliminating the biases unaffected by the number of
simulations we did not use any variance reduction technique that
can be applied to reduce the statistical error of the estimates,
see e.g. Boyle, Broadie and Glassermann (1997).

The comparison of the Monte Carlo estimates with the exact option
prices calculated by analytical formulae demonstrates that $Q$ is
an unbiased estimator (the exact value is inside the 0.95
confidence interval of the estimates for any $M)$. The discretely
monitored barrier option estimate $Q_S $ converges to the
continuous barrier case as $M$ increases. However, the convergence
is very slow and the bias is larger than 1{\%} of the true price
even for 1024 time steps. The use of the unbiased estimator
(\ref{eq8}) in the above examples is very efficient because one
time step is enough to obtain the unbiased option price estimate
while the standard discrete-time approach (\ref{eq5}) requires an
enormous number of time steps.

\section{Biased estimators for multi-barrier option via
Fr\'{e}chet bounds}

\subsection{Fr\'{e}chet bounds for distribution of
Brownian Bridge extrema}

The joint distribution $P^{(m)}$ in (\ref{eq9}) can be found in
closed form via infinite series for the case of two dependent
extrema (that is two barriers in the same time region) using the
results from Andersen (1998) for double barrier on a single asset
and He, Keirstead and Rebholz (1998) for two barriers imposed on
different assets. Then, in principle, the unbiased option price
estimator can be calculated using (\ref{eq8}). In general, the
joint distribution, $P^{(m)}$, of three and more dependent extrema
(that is three and more barriers in the same time region) is
unknown and unbiased option price estimator (\ref{eq8}) can not be
calculated. However, the univariate marginal distributions of the
extremes, given by (\ref{eq10}), are known and very simple. The
classes of multivariate distributions with given margins are the
so-called Fr\'{e}chet classes. The results on bounds of the
distributions with known margins and unknown dependence structure
can be found in multivariate distribution theory, see for example
Joe (1997). The upper and lower bounds, so-called Fr\'{e}chet
bounds, for the unknown joint distribution with known univariate
margins are based on simple inequalities involving probabilities
of sets.

\textbf{Theorem (}Lemma 3.8 in Joe (1997)) \textit{Let }$A_1 ,...,A_k $\textit{ be the events such that }$\mbox{Pr}(A_i ) =
a_i , \quad \;i = 1,...,k$\textit{. Then }

\begin{equation}
\label{eq12}
\max [0,\sum\limits_{i = 1}^k {a_i } - (k - 1)] \le \mbox{Pr}(A_1 \cap ...
\cap A_k ) \le \mathop {\min }\limits_{i = 1,...,k} a_i .
\end{equation}

Let $B_i = \{L_i^{(m)} > h_i^{(m)} \}$ and $A_i = \{M_i^{(m)} < H_i^{(m)}
\}$ be the events that minimum is above the lower barrier and maximum is
below the upper barrier for the $i$-th asset on $[t_m ,t_{m + 1} ]$
respectively. Then $\mbox{Pr}(B_i ) = [1 - \xi _i^{(m)} (h_i^{(m)} )]$,
$\mbox{Pr}(A_i ) = [1 - \xi _i^{(m)} (H_i^{(m)} )]$ and the above theorem
gives the following bounds for the joint probability $P^{(m)}$ in (\ref{eq9})

\begin{equation}
\label{eq13}
\begin{array}{l}
 P^{(m)} \ge P_L^{(m)} = \max [1 - \sum\limits_{i = 1}^d {[\xi _i^{(m)}
(H_i^{(m)} )} + \xi _i^{(m)} (h_i^{(m)} )],\;0], \\
 P^{(m)} \le P_U^{(m)} = \mathop {\min }\limits_{i = 1,...,d} [1 - \xi
_i^{(m)} (H_i^{(m)} ),\;1 - \xi _i^{(m)} (h_i^{(m)} )]. \\
 \end{array}
\end{equation}

\noindent The upper bound $P_U^{(m)} $ corresponds to a perfect
positive dependence between all events $A_i ,\;B_i $, $i =
1,...,d$. If there are only two events then the lower bound
$P_L^{(m)} $ corresponds to a perfect negative dependence between
the events. Perfect positive (negative) dependence between $A_i $
and $A_j $, $B_i $ and $B_j $ means a perfect positive (negative)
dependence between $M_i^{(m)} $ and $M_j^{(m)} $, $L_i^{(m)} $ and
$L_j^{(m)} $ respectively. Perfect positive (negative) dependence
between $A_i $ and $B_j $ means a perfect negative (positive)
dependence between $M_i^{(m)} $ and $L_j^{(m)} $\footnote{The
random variables $X$ and $Y$ with continuous marginal
distributions have perfect positive (negative) dependence if
$X=T(Y)$ where $T$(.) is a strictly increasing (decreasing)
function.}. Also, consider the joint probability of the
independent events\footnote{ This joint distribution has been used
by Andersen (1998) and Beaglehole, Dybvig and Zhou (1997) to
estimate double barrier and double lookback options on a single
asset.} $A_i ,\;B_i $, $i = 1,...,d$

\begin{equation}
\label{eq14}
P_I^{(m)} = \prod\limits_{i = 1}^d {[1 - \xi _i^{(m)} (H_i^{(m)} )]} \cdot
[1 - \xi _i^{(m)} (h_i^{(m)} )],
\end{equation}

\noindent
which, of course, satisfy $P_L^{(m)} \le P_I^{(m)} \le P_U^{(m)} $. If all
events are positively (negatively) dependent then $P_I^{(m)} \le P^{(m)}$
($P_I^{(m)} \ge P^{(m)})$.

\subsection{Fr\'{e}chet bounds and method of images}

The Fr\'{e}chet bounds (\ref{eq13}) for the joint distribution $P^{(m)}$ in (\ref{eq9}) have
the following simple interpretation via the method of images. The joint
distribution can be obtained from a solution of the Kolmogorov forward
equation with absorbing boundaries using (\ref{eq9}). The solution for the case of a
single barrier imposed on the $k$-th asset in $[t_m ,t_{m + 1} ]$ can be found
by the method of images, see for example Cox and Miller (1965). The method
is based on finding the linear combination of the unrestricted process
solutions $p_0 (\vec {S}^{(m + 1)}\vert \vec {S}^{(m)})$ started at $\vec
{S}^{(m)}$, the so-called source, and $p_0 (\vec {S}^{(m + 1)}\vert \vec
{X}^{(m)})$ started at $\vec {X}^{(m)}$, the so-called primary
image\footnote{ In the case of few barriers the primary image may create
further images.}, satisfying the initial and absorbing boundary conditions.
The location of the primary image after \textit{log}-scale change is found by reflection
of the source in respect to the boundary. This leads to the formula (\ref{eq10}),
where the contribution of the source is represented by 1 and the
contribution of the primary image is represented by $ - \xi _k^{(m)}
(h_k^{(m)} )$ (or $ - \xi _k^{(m)} (H_k^{(m)} ))$ for the case of lower (or
upper) barrier. Now it is easy to see that the lower bound $P_L^{(m)} $ in
(\ref{eq13}) is obtained from the source and all its primary images (one image for
each barrier). The upper bound $P_U^{(m)} $ in (\ref{eq13}) is obtained from the
source and one of the primary images that gives the largest contribution.
The method of images cannot be used to find the exact transition probability
in the general case where few barriers are imposed on different assets with
arbitrary correlation. However, formally, primary images can always be
introduced and used to get approximate solutions. The numerical results
below will demonstrate that these approximations are very effective.

\subsection{Three biased estimators for multi-barrier option}

Using the estimators $P_X^{(m)} ,\;X = L,I,U$ defined in (\ref{eq13}) and (\ref{eq14}) for
the joint probability of the extremes (\ref{eq9}) we can form three biased
estimators for the option price (\ref{eq8}):

\begin{equation}
\label{eq15}
Q_X = V(\vec {S}(T)) \cdot \mbox{{\rm I}}_{\tau > T} \prod\limits_{m = 1}^M
{P_X^{(m)} } ,\;X = U,I,L.
\end{equation}

\noindent
It is easy to see from (\ref{eq8}), (\ref{eq13}) and
(\ref{eq14}) that

\begin{equation}
\label{eq16}
Q_L \le Q \le Q_U ,\;\;Q_L \le Q_I \le Q_U
\end{equation}

\noindent
and, of course, $Q_X \le Q_S $ because $P_X^{(m)} \le 1,\;X = U,I,L$. While
$Q_I $ is typically the most accurate, $Q_L $ and $Q_U $ are more useful
because they can always be used to bind the true value by the confidence
interval

\begin{equation}
\label{eq17}
[Q_L - z_{1 - \alpha / 2} \textstyle{{s(Q_L )} \over {\sqrt N }},Q_U + z_{1
- \alpha / 2} \textstyle{{s(Q_U )} \over {\sqrt N }}],
\end{equation}

\noindent
where $z_{1 - \alpha / 2} $ is a quantile of the standard Normal
distribution, $s(.)$ is the standard deviation, $N$ is the number of simulated
paths and $\alpha $ is the significance level. Then the true value lies
within the interval with at least $1 - \alpha $ probability. The middle of
the interval,

\begin{equation}
\label{eq18}
Q_0 \approx \textstyle{1 \over 2}(Q_L + Q_U )
\end{equation}

\noindent
can be used as a point estimator for the true value and half of the interval
with $z_{1 - \alpha / 2} $=1 can be called the standard error of the
estimator. Using the inequality $P_I^{(m)} \ge P^{(m)}$ ($P_I^{(m)} \le
P^{(m)})$ for negatively (positively) dependent events it is easy to find,
for some simple barrier and dependence structures, that $Q_I $ gives better
upper or lower estimator (however, in general, we do not know this $a$
\textit{priori}). Then the true option price can be estimated by

\begin{equation}
\label{eq19}
Q_1 \approx \textstyle{1 \over 2}(Q_L + Q_I )\quad \mbox{or}\quad Q_2
\approx \textstyle{1 \over 2}(Q_I + Q_U )
\end{equation}

\noindent
respectively. The confidence intervals to bind the true value in these cases
are formed by analogy with (\ref{eq17}). Here we list some examples. If the barrier
structure consists of two barriers (upper and lower) imposed on one of the
assets in each time region then $Q_I \ge Q$ because the events of the upper
and lower barrier hits are always negatively dependent (that is maximum and
minimum of the asset are always positively dependent). If the barrier
structure consists of two upper or two lower barriers imposed on two
positively (negatively) correlated assets in each time region then $Q_I \le
Q$ ($Q_I \ge Q)$ because the events of the barrier hits are positively
(negatively) dependent. Also, $Q_I \le Q$ if only lower or only upper
barriers are imposed on positively correlated assets in each time region.

All three option price biased estimators $Q_X ,\;X = U,I,L$ given by (\ref{eq15})
are trivial to implement because the payoff weights $\prod\limits_{m = 1}^M
{P_X^{(m)} } ,\;X = U,I,L$ are based on simple marginal distributions (\ref{eq10}).
As $M$ increases, it becomes less and less likely that hits of the different
barriers will occur within the same time interval and the events of barrier
hits (and corresponding extremes) become disjointed. That is, the
probability distribution $P^{(m)}$ for each of the time regions $[t_m ,t_{m
+ 1} ]$ is one of the univariate marginal distributions (\ref{eq10}) in the limit
$t_{m + 1} - t_m \to 0$. Disjoining of the extrema means that the option
price estimator becomes independent from the dependence structure (the
so-called copula) of the extrema and depends on their marginal distributions
only\footnote{ The phenomenon of maximum and minimum ``decoupling'' has been
noted by Andersen (1998) while using independently drawn maximum and minimum
to estimate double barrier and double lookback options on a single asset.}.
It implies that $Q_I $, $Q_L $ and $Q_U $ should converge to the true value.
Convergence of $Q_I $ and $Q_U $ to each other can be used as a weak
criterion of the extreme disjoining. Also note that the standard
discrete-time biased estimator, $Q_S $, given by (\ref{eq5}), always converges to
the true value and $Q_S > Q_U \ge Q$. In the next Section we will
demonstrate numerically that all three estimators $Q_X ,\;X = U,I,L$ are
rapidly convergent to the true value when compared to $Q_S $. The rate of
convergence depends on the barrier and asset correlation structures. In this
paper we do not pursue the analytical derivation of the convergence rate but
find it numerically for some basic cases. The numerical examples presented
in the following Section demonstrate the effectiveness of the estimators
(\ref{eq15}) to correctly price barrier options.

\section{Performance of the biased estimators}
\label{sec:performance}

To demonstrate the performance (rapid convergence) of the biased estimators
$Q_X ,\;X = U,I,L$ given in (\ref{eq15}) we calculate these estimators and the
standard discrete-time estimator $Q_S $, see (\ref{eq5}), versus the number of
equally spaced sampling dates $M$ (the time step is $\delta t = T / M)$ for the
cases of options with two or more knock-out barriers. In addition, we show
the results for the point estimator $Q_0 \approx (Q_L + Q_U ) / 2$, see
(\ref{eq18}). We assume that all parameters (volatilities, interest rate, barriers,
correlations) are constant and there are no continuous dividends. Being
interested in the convergence of the biases unaffected by the number of
simulations we did not use any variance reduction techniques that can be
applied to reduce the statistical error of the estimates. In all examples
the discounted payoff (which is paid at maturity if the assets never hit the
barriers) is always determined by the first asset: $V(S_1 (T)) = e^{ -
rT}\max [S_1 (T) - K,0]$. The convergence rate should be irrelevant to the
payoff paid at maturity.

\subsection{Double knock-out call on a single asset}

First we consider the double knock-out call on a single asset with the
lower, $h_1 $, and upper, $H_1 $, barriers. The explicit expressions for the
probability bounds (\ref{eq13}), (\ref{eq14}) used in the biased estimators (\ref{eq15}) are

\[
\begin{array}{l}
 P_U^{(m)} = \min [1 - \xi _1^{(m)} (h_1 ),\;1 - \xi _1^{(m)} (H_1 )],\; \\
 P_I^{(m)} = [1 - \xi _1^{(m)} (h_1 )] \cdot [1 - \xi _1^{(m)} (H_1 )],\; \\
 P_L^{(m)} = \max [1 - \xi _1^{(m)} (h_1 ) - \xi _1^{(m)} (H_1 ),\;0]. \\
 \end{array}
\]

\noindent
Actually, the exact joint distribution $P^{(m)}$ is
known. It is represented by infinite series (usually the series
are rapidly convergent) and can be used to calculate the unbiased
option price estimator, see Andersen (1998). Being mainly
interested in the convergence of the biased estimators we do not
pursue this calculation. In Table 2 we show the performance of the
tested estimators. The maximum and minimum of the asset on the
same time interval are always positively dependent (the hits of
the lower and upper barriers are negatively dependent) thus the
estimator $Q_I $ based on the distribution of the independent
extremes is always larger than the unbiased estimator $Q$. In this
case $Q_I $ is a better upper estimator than $Q_U $. We present
the results for $Q_1 \approx (Q_I + Q_L ) / 2$, see (\ref{eq19}),
which can be used as a better point estimator for the true price
instead of $Q_0 \approx (Q_U + Q_L ) / 2$. All our biased
estimators $Q_U $, $Q_I $ and $Q_L $ are rapidly convergent to the
true value in comparison to the standard estimator $Q_S $.

\noindent Comparison of our biased estimators with the exact
analytical result shows that the biases of $Q_I $, $Q_L $ and $Q_U
$ are less than their standard errors for $M \ge 4$, $M \ge 4$ and
$M \ge 8$ respectively while the bias of the standard estimator
$Q_S $ is significantly larger than its standard error even for
$M$=1024. The standard errors are less then 1{\%} of the true
value. The exact value is always inside the standard confidence
intervals of the point estimators $Q_1 $ and $Q_0 $. In the case
where a double barrier is imposed on the asset the hits of the
upper and lower barriers are ``physically'' distant (the asset can
not be close to the upper and lower barriers at the same time).
Thus intuitively we expect that the bias $Q_U - Q_L $ should be
exponentially small for large $M$. In Figure 1 we plot $\ln (Q_U -
Q_L )$ versus $M$ for the case of the double knock-out call
considered in Table 2. The standard error of the plotted estimates
is less than the size of the symbols. The observed linear
behaviour of the graph indicates that the bias decreases as $Q_U -
Q_L \sim e^{ - \alpha / \delta t}$.

\subsection{Two asset call with two knock-out barriers}

To demonstrate convergence of the estimators for the case where barriers are
imposed on different arbitrary correlated assets we consider two asset
down-and-out call with the lower barriers $h_1 $ and $h_2 $ imposed on the
first and second assets respectively. In this case the explicit expressions
for the probability bounds (\ref{eq13}), (\ref{eq14}) used in the biased estimators (\ref{eq15}) are

\begin{equation}
\label{eq20}
\begin{array}{l}
 P_U^{(m)} = \mathop {\min }\limits_{i = 1,..,d} [1 - \xi _i^{(m)} (h_i
)],\; \\
 P_I^{(m)} = \prod\limits_{i = 1}^d {[1 - \xi _i^{(m)} (h_i )]} ,\; \\
 P_L^{(m)} = \max [1 - \sum\limits_{i = 1}^d {\xi _i^{(m)} (h_i )} ,0], \\
 \end{array}
\end{equation}

\noindent where $d$ = 2. We designed the problem parameters to
equate probabilities of the barrier hits $\xi _1^{(m)} (h_1 ) =
\xi _2^{(m)} (h_2 )$ if $\rho = \;1$. In this case $P_L^{(m)} =
\max [1 - 2\xi _1^{(m)} (h_1 ),0]$, $P_U^{(m)} = 1 - \xi _1^{(m)}
(h_1 )$ and we expect worst convergence because the events of the
barrier hits do not become disjoint at $M \to \infty $. The exact
joint distribution $P^{(m)}$ can be found in a closed form using
the results in He, Keirstead and Rebholz (1998). It is expressed
via infinite series (usually the series are rapidly convergent)
and can be used to calculate the unbiased estimator (\ref{eq8}).
Again, being only interested in the convergence of the biased
estimators we do not pursue this calculation.

In Table 3 we show the exact prices and Monte Carlo estimators for
various asset correlation values. The exact option price for $\rho
= - 0.5,\;0.5$ has been found via numerical integration of the
two-dimensional density function from He, Keirstead and Rebholz
(1998). In the cases $\rho = - 1,\;0,\;1$ the option value can be
expressed via the barrier options on a single asset and the exact
solutions are represented via the standard cumulative Normal
function\footnote{ For $\rho = - 1$ the option is reduced to a
double knock-out call on the first asset with the exponentially
growing upper and flat lower barriers. For $\rho = 0$ the option
can be represented as a product of a down-and-out call on the
first asset and a down-and-out digital option on the second asset.
For $\rho = 1$ the option is reduced to a down-and-out call on the
first asset.}. If the correlation, $\rho $, between the assets is
positive (negative) then the minima of the assets over the same
time interval are always positively (negatively) dependent. Thus
the estimator $Q_I $ based on the distribution of the independent
extremes is always larger (less) than the unbiased estimator $Q$
if $\rho < 0$ ($\rho > 0)$. We do not present the results for $Q_1
\approx (Q_I + Q_L ) / 2$ (if $\rho < 0)$ and $Q_2 \approx (Q_I +
Q_U ) / 2$ (if $\rho > 0)$ but they can be used as a better point
estimate for the true price instead of $Q_0 \approx (Q_U + Q_L ) /
2$. If $\rho = 0$ then the minima are independent and $Q_I $ is an
unbiased estimator for the true price because $P_I^{(m)} $ is a
valid joint distribution. If $\rho = 1$ then the minima of the
assets have a perfect positive dependence and $Q_U $ is an
unbiased estimator because $P_U^{(m)} $ is a valid joint
distribution.

The obtained results show that our biased estimators $Q_U $, $Q_I $ and $Q_L
$ are rapidly convergent to the true value when compared to the standard
estimator $Q_S $. Comparison of our results with the exact results shows
that the biases of $Q_I $, $Q_L $ and $Q_U $ are less then their standard
errors if $M \ge 16$ while the bias of the standard estimator $Q_S $ is
significantly larger than its standard error even for $M$=1024 (the standard
errors are less then 1{\%} of the true value for almost all cases). The only
case when the convergence of $Q_I $ and $Q_L $ is not so rapid is $\rho = 1$
(in this case $Q_U $ is the unbiased estimator). The standard confidence
interval of the point estimator $Q_0 $ always contains the exact price.

The convergence rate of our biased estimators strongly depends on
correlation between the assets. In Figure 2, Figure 3 and Figure 4
we present the graphs indicating the convergence rates for $\rho =
- 1,\;0,\;1$. The size of the symbols used for the graphs is
larger than the standard error of the estimates (we have used more
simulations for estimates at large $M)$.

The linear behaviour of the graphs at large $M$ indicates the
following. If $\rho = - 1$ then $Q_U - Q_L \sim e^{ - \beta /
\delta t}$ for $M > > 1$. This is similar to the results for
double barrier call in Table 2 and Figure 1, because the assets
have a perfect negative dependence and the option is equivalent to
a single asset knock-out option with the flat lower and
exponentially growing upper barriers . If $\rho = 1$ then $Q_U -
Q_L \sim \sqrt {\delta t} $ for $M > > 1$. This square root
convergence is the worst observed rate. In this case the asset
minima have a perfect positive dependence and the estimator $Q_U $
is the unbiased estimator of the true option price. That is $Q_L $
and $Q_S $ converge to the true price at the same rate from below
and above respectively. If $\rho = 0$ then $Q_U - Q_L \sim \delta
t^2$ for $M > > 1$. We have observed that the convergence rate
smoothly deteriorates from the best exponential decay $\sim \exp (
- \beta / \delta t)$ at $\rho = - 1$ to a rapid power decay
$\delta t^2$ at $\rho = 0$ and slow square root decreasing $\sim
\sqrt {\delta t} $ at $\rho = 1$ as the correlation is changed
between $ - 1$ and $1$. Note that we have designed the parameters
to get worst convergence at $\rho = 1$. If we change the
parameters to make the barrier hit probabilities unequal then the
extrema should become disjointed as $M \to \infty $ even for $\rho
= 1$. For example, if we set the asset spots $S_1 (0) = 95$, $S_2
(0) = 105$\textbf{ }and do not change the other parameters then
the convergence rate at $\rho = 1$ is rapid exponential decay
$\sim \exp ( - \beta / \delta t)$, see Figure 5, while for $\rho =
0$ and $ - 1$ the rates do not change (that is $\sim \delta t^2$
and $\sim \exp ( - \beta / \delta t)$ respectively).

\subsection{Multi-asset call with multiple barriers}

Finally, to show that our biased estimators work well for real
multi-dimensional problems we consider, see Table 4, down-and-out
call on $d$ underlying assets with the lower barriers imposed on
all assets for the cases $d = 3$ and $d = 10$. The exact
(analytical) result is not available for this problem. Explicit
expressions for the probability bounds (\ref{eq13}), (\ref{eq14})
used in the biased estimators (\ref{eq15}) are given by
(\ref{eq20}). As we have chosen positive correlation between all
assets the asset minima are positively dependent. Thus $Q_I $ is
always less than the unbiased estimator $Q$ and we present the
results for $Q_2 \approx (Q_I + Q_U ) / 2$ which is a better point
estimate for the true price than $Q_0 $. As in the previous
examples, all our biased estimators $Q_U $, $Q_I $ and $Q_L $ are
rapidly convergent to each other. Their biases become less than
the statistical errors for $M \ge 16$ while the bias of the
standard estimator $Q_S $ is larger than its standard error even
for $M = 1024$.

\section{Conclusions and discussion}

In this paper we have developed a conditioning technique, that can be called
a Brownian Bridge scheme, for Monte Carlo simulation of a general class of
multi-asset options with continuously monitored knock-out barriers imposed
for some or all underlying assets. We have derived the general formula (\ref{eq8})
for an unbiased estimator of the option based on the joint distribution of
the multi-dimensional Brownian Bridge extrema (\ref{eq9}). If the distribution is
known, for example, one barrier imposed on one of the assets (or two
barriers imposed on different or the same assets) per time region, the
scheme provides a simple unbiased estimator. The barriers, drifts and
volatilities are required to be piecewise constant functions of time. In the
case of more than two barriers per time region the distribution is unknown
in a closed form for arbitrary dependence between the assets and we derived
the upper, $Q_U $, and lower, $Q_L $, biased option price estimators. The
estimators are based on the Fr\'{e}chet lower, $P_L^{(m)} $, and upper,
$P_U^{(m)} $, bounds (\ref{eq13}) for the unknown joint distribution with given
univariate margins. We have also used the estimator $Q_I $ based on the
joint distribution of independent extrema, $P_I^{(m)} $. For some simple
barrier and dependence structures $Q_I $ can provide better upper or lower
bounds. While this estimator is usually more accurate, $Q_L $ and $Q_U $ are
often more useful because they can always be used to bind the true option
price by the confidence interval. As the time between the sampling dates
decreases, the Brownian Bridge extrema become more disjointed and the biased
estimators $Q_L $, $Q_I $, $Q_U $ converge to each other and the true value.
In the limit, the option price estimator becomes independent of the
dependence structure (the so-called copula) of the extrema and depends on
their marginal distributions only. In our numerical examples we showed that
the bias $Q_U - Q_L $ is less and convergence is rapid when compared to the
bias and convergence of the standard estimator $Q_S $ ($Q_S - Q\sim \sqrt
{\delta t} $ for $M > > 1$, where $\delta t = T / M)$. Usually, $Q_U - Q_L $
is less than the statistical error of the estimates for only a few time
steps. In practice, the intermediate dates are introduced due to the
interest rate or volatility term-structures and the insertion of the
additional sampling dates to eliminate the bias may not be even required.
The convergence rate depends on the barrier and correlation structures. We
have always observed $Q_U - Q_L \sim \exp ( - \alpha / \delta t)$ for the
case of lower and upper barriers imposed on the same asset. For the case of
two lower barriers imposed on two assets the best detected convergence rate
is exponential decay $\sim \exp ( - \beta / \delta t)$ and the worst
detected rate is slow square root decreasing $\sim \sqrt {\delta t} $. The
worst case was obtained for the unrealistic special set of parameters ($\rho
= 1$ and identical parameters for both assets) which makes barrier hits
always equal each other and $Q_U $ to be an unbiased estimator of the true
price.

The described Brownian Bridge technique is straightforward to use for the
valuation of the knock-in barrier options and barrier options with constant
rebates, $R$, paid at maturity. The unbiased estimators in these cases are
given by

$$
Q = V(S(T))[1 - {\rm I}_{\tau > T} \prod\limits_{m = 1}^M
{P^{(m)}} ]
$$

and

$$
Q = V(S(T)){\rm I}_{\tau > T} \prod\limits_{m = 1}^M {P^{(m)}} + R
\cdot [1 - {\rm I}_{\tau > T} \prod\limits_{m = 1}^M {P^{(m)}} ]
$$

\noindent
respectively. The upper and lower biased estimators can easily be calculated
using $P_L^{(m)} $ and $P_U^{(m)} $. The technique can easily be applied to
efficiently estimate discretely monitored barrier options with a large
number of observation dates using the method proposed by Andersen (1996).
That is, using our scheme calculate the continuously monitored barrier
option price $Q_c $ and using the standard Monte Carlo method estimate the
option with a low frequency monitored barrier. Then the interpolation
formula $Q_M \approx Q_c + \lambda / \sqrt M $, where $Q_M $ is a barrier
option with $M$ monitored dates, allows for effective estimation of the
option with a high frequency monitored barrier.

Finding the upper and lower biased estimators is not
straightforward for the case of multi-barrier options with rebates
paid at hitting times. This type of problem is also relevant to
the valuation of credit derivatives. To calculate the unbiased
option price estimator, multiple hitting times should be simulated
from their valid joint distribution which is not known even for
the case of two barriers. However, the hitting times can easily be
simulated from their known univariate marginal distributions, see
Anderson (1996). Thus, again we have a problem of the unknown
joint distribution with the known univariate margins. Such options
can be evaluated in the following way. Calculate one option
estimate $Q^{(\ref{eq1})}$ simulating the hitting times marginally
with perfect positive dependence\footnote{ The random variables
$X_i ,\;i = 1,...,k$ with continuous marginal distributions $F_i $
and perfect positive dependence can be simulated as $X_i = F_i^{ -
1} (U)$, where $U$ is a random variable from $U(0,1)$. If $X_i
,\;i = 1,2$ have a perfect negative dependence then $X_1 = F_1^{ -
1} (U)$, $X_2 = F_2^{ - 1} (1 - U)$. If $X_i ,\;i = 1,...,k$ are
independent then $X_i = F_i^{ - 1} (U_i )$, where $U_i ,\;i =
1,...,k$ are independent random variables from $U(0,1)$.}. Another
estimate $Q^{(\ref{eq2})}$ can be found by simulating the events
marginally with perfect negative dependence in the case of two
barriers or simulating the events independently if there are more
than two barriers. The difference between $Q^{(\ref{eq1})}$ and
$Q^{(\ref{eq2})}$ can be used as a weak criterion for the
disjointed events to justify their marginal simulations. That is,
if the difference is less than the statistical error then the
events can be assumed disjointed enough (marginal simulation is
justified) and $(Q^{(\ref{eq1})} + Q^{(\ref{eq2})}) / 2$ can be
used as a point estimate for the true price. Otherwise additional
sampling dates need to be inserted. Lookback type options with
payoff dependent on a few continuous extrema can be evaluated in a
similar way. We will consider these problems in further research.

In all our examples we have assumed the lognormal diffusion process (\ref{eq1}) to
allow for comparison with analytical results. However, the Brownian Bridge
scheme discussed here is still applicable for more general diffusion
processes. For example, it is applicable if drifts and volatilities are
state-dependent. To solve these models approximate discretisation schemes
freezing the drifts and volatilities to the left point of each simulation
step are used, see for example Kloeden and Platen (1992). Then Brownian
Bridge schemes can be used because the requirements of piecewise constant
drifts and volatilities are satisfied. This will eliminate the bias due to
discrete underestimation of the continuous extrema (bias due to the
discretisation scheme will not be removed).

In this paper we have focused on the application of the technique for the
valuation of simple knock-out multi-asset options. However, practical use of
the technique lies in a broad range of problems. For example, the technique
can potentially be used for credit and market risk problems where valuation
of a multi-asset payoff with some barrier levels imposed on the underlying
assets is very essential.

\begin{center}
\textbf{Acknowledgements}
\end{center}

\noindent
The author thanks Volf Frishling and Frank de Hoog for
stimulating discussions and helpful advice.

\baselineskip=12pt

\begin{center}
\textbf{References}
\end{center}

\begin{description}
\item
{Andersen, L., and Brotherton-Racliffe, R. (1996). Exact Exotics.
{\it Risk}, {\bf 9}(10), 85-89.}

\item{
Andersen, L. (1996). Monte Carlo simulation of Barrier and
Lookback Options with continuous or High-Frequency Monitoring of
the Underlying Asset. Working paper, General Re Financial
Products.}

\item
{Andersen, L. (1998). Monte Carlo simulation of Options on Joint
Minima and Maxima. Working paper, General Re Financial Products.}

\item
Beaglehole, D. R., Dybvig, P. H., and Zhou, G. (1997). Going to
extremes: Correcting Simulation Bias in Exotic Option Valuation.
{\it Financial Analyst Journal}, January/February, 62-68.

\item
Borodin, A., and Salminen, P. (1996). {\it Handbook of Brownian
Motion-Facts and Formulae}. Birkhauser Verlag, Basel.

\item
Boyle, P., Broadie, M., and Glassermann, P. (1997). Monte Carlo
Methods for Security Pricing. {\it Journal of Economic Dynamics
and Control}, {\bf 21}, 1267-1321.

\item
Broadie, M., Glasserman, P., and Kou, S. (1997). A continuity
correction for discrete barrier options. {\it Mathematical
Finance}, {\bf 7}, 325-349.

\item
Cox, D., and Miller, H. (1965). {\it The Theory of Stochastic
Processes}. Chapman and Hall.

\item
Dewynne, J., and Wilmott, P. (1994). Partial to Exotic. {\it
Risk}, December, 53-57.

\item
He, H., Keirstead, W. P., and Rebholz, J. (1998). Double Lookback.
{\it Mathematical Finance}, {\bf 8}(3), 201-228.

\item
Heynen, R., and Kat, H. (1994a). Partial Barrier Options. {\it
Journal of Financial Engineering}, {\bf 3}(3/4), 253-274.

\item
Heynen, R., and Kat, H. (1994b). Crossing barriers. {\it Risk},
{\bf 7}(6), 46-51.

\item
Hull, J., and White, A. (1993). Efficient Procedures for Valuing
European and American Path-Dependent Options. {\it Journal of
Derivatives}, Fall, 21-31.

\item
Joe, H. (1997). Multivariate Models and Dependence Concepts.
Chapman {\&} Hall.

\item
Kat, H., and Verdonk, L. (1995). Tree Surgery. {\it Risk},
February, 53-56.

\item
Karatzas, I., and Shreve, S. (1991). {\it Brownian Motion and
Stochastic Calculus}. Springer Verlag.

\item
Kloeden, P., and Platen, E. (1992). {\it Numerical Solution of
Stochastic Differential Equations}. Springer Verlag.

\item
Kunitomo, N., and Ikeda, M. (1992). Pricing Options with Curved
Boundaries. {\it Mathematical Finance}, {\bf 4}, 275-298.

\item
Merton, R. (1973). Theory of Rational Pricing. {\it Bell Journal
of Economics and Management Science}, {\bf 4}, Spring, 141-183.

\item
Rubinstein, M., and Reiner, E. (1991). Breaking down the barriers.
{\it Risk}, {\bf 4}(8), 28-35.

\end{description}

\begin{table}[p]
\begin{center}
\begin{tabular}
{|p{35pt}|p{72pt}|p{72pt}|p{35pt}|p{72pt}|p{72pt}|} \hline
\multicolumn{3}{|p{170pt}|}{ One asset down and out call.
\newline The exact value is 8.794. } &
\multicolumn{3}{|p{179pt}|}{Two asset down-and-out call. \newline The exact value is 8.256.}  \\
\hline $M$& $Q$(std.err.)& $Q_S $(std.err.)& $M$& $Q$(std.err.)&
$Q_S $(std.err.) \\
\hline 1 \par 2 \par 4 \par 16 \par 64 \par 256 \par 1024&
8.79(0.02) \par 8.80(0.02) \par 8.80(0.02) \par 8.79(0.02) \par
8.80(0.02) \par 8.80(0.02) \par 8.80(0.02)& 10.91(0.02) \par
10.66(0.02) \par 10.32(0.02) \par 9.74(0.02) \par 9.33(0.02) \par
9.08(0.02) \par 8.94(0.02)& 1 \par 2 \par 4 \par 16 \par 64 \par
256 \par 1024& 8.26(0.02) \par 8.26(0.02) \par 8.27(0.02) \par
8.27(0.02) \par 8.28(0.02) \par 8.28(0.02) \par 8.28(0.02)&
14.93(0.03) \par 13.62(0.03) \par 12.35(0.03) \par 10.52(0.02) \par 9.47(0.02) \par 8.90(0.02) \par 8.59(0.02) \\
\hline
\end{tabular}
\caption{One asset down and out call: $S_1 (0) = 100$, $K = 100$,
$h_1 = 90$, $\sigma _1 = 0.3$, $r = 0.1$, $T = 0.5$, $400\,000$
simulations. Two asset down-and-out call with a single barrier:
$S_1 (0) = S_2 (0) = 100$, $K = 100$, $h_2 = 90$, $r = 0.1$, $T =
1$, $\sigma _1 = \sigma _2 = 0.3$, $\rho = 0.5$, $800\,000$
simulations.} \label{tab1}
\end{center}
\end{table}

\begin{table}[p]
\begin{center}
\begin{tabular}
{|p{31pt}|p{61pt}|p{61pt}|p{61pt}|p{61pt}|p{61pt}|p{61pt}|} \hline
$M$ & $Q_U $(std.err.)& $Q_I $(std.err.)& $Q_L $(std.err.)& $Q_S
$(std.err.)& $Q_1 $(std.err.)&
$Q_0 $(std.err.) \\
\hline 1 \par 2 \par 4 \par 8 \par 16 \par 64 \par 256 \par 1024&
3.01(0.01) \par 2.21(0.01) \par 1.84(0.01) \par 1.79(0.01) \par
1.78(0.01) \par 1.78(0.01) \par 1.78(0.01) \par 1.78(0.01)&
2.41(0.01) \par 1.89(0.01) \par 1.79(0.01) \par 1.79(0.01) \par
1.78(0.01) \par 1.78(0.01) \par 1.78(0.01) \par 1.78(0.01)&
1.11(0.01) \par 1.72(0.01) \par 1.78(0.01) \par 1.79(0.01) \par
1.78(0.01) \par 1.78(0.01) \par 1.78(0.01) \par 1.78(0.01)&
12.23(0.04) \par 9.60(0.04) \par 7.41(0.03) \par 5.73(0.03) \par
4.50(0.02) \par 3.06(0.02) \par 2.40(0.02) \par 2.08(0.02)&
1.76(0.66) \par 1.80(0.09) \par 1.79(0.01) \par 1.79(0.01) \par
1.78(0.01) \par 1.78(0.01) \par 1.78(0.01) \par 1.78(0.01)&
2.06(0.96) \par 1.97(0.26) \par 1.81(0.04) \par 1.79(0.01) \par 1.78(0.01) \par 1.78(0.01) \par 1.78(0.01) \par 1.78(0.01) \\
\hline
\end{tabular}
\caption{Double knock-out call on a single asset. $h_1 = 900$,
$H_1 = 1100$, $S(0) = K = 1000$, $\sigma _1 = 0.2$, $r = 0.1$, $T
= 0.5$, $400\,000$ simulations. The exact value of the
continuously monitored option is 1.793.}

\label{tab2}
\end{center}
\end{table}

\begin{table}[p]
\begin{center}
\begin{tabular}
{|p{20pt}|p{26pt}|p{63pt}|p{63pt}|p{63pt}|p{63pt}|p{63pt}|p{35pt}|}
\hline $\rho $ & $M$ & $Q_U $(std.err.)& $Q_I $(std.err.)& $Q_L
$(std.err.)& $Q_S $(std.err.)& $Q_0 $(std.err.)&
$Q_c $ \\
\hline 0& 1 \par 8 \par 16 \par 32 \par 64 \par 1024& 5.02(0.03)
\par 3.78(0.04) \par 3.70(0.04) \par 3.66(0.04) \par 3.65(0.04)
\par 3.64(0.04)& 3.65(0.03) \par 3.66(0.04) \par 3.66(0.04) \par
3.65(0.04) \par 3.65(0.04) \par 3.64(0.04)& 2.27(0.02) \par
3.62(0.04) \par 3.65(0.04) \par 3.65(0.04) \par 3.65(0.04) \par
3.64(0.04)& 11.76(0.07) \par 6.84(0.06) \par 5.92(0.05) \par
5.27(0.05) \par 4.81(0.05) \par 3.93(0.05)& 3.64(1.41) \par
3.70(0.12) \par 3.67(0.06) \par 3.65(0.05) \par 3.65(0.04) \par
3.64(0.04)&
3.649 \\
\hline 0.5& 1 \par 8 \par 16 \par 32 \par 64 \par 1024& 7.78(0.05)
\par 6.71(0.06) \par 6.61(0.06) \par 6.57(0.06) \par 6.55(0.06)
\par 6.54(0.06)& 5.84(0.04) \par 6.48(0.05) \par 6.53(0.06) \par
6.54(0.06) \par 6.54(0.06) \par 6.54(0.06)& 4.22(0.04) \par
6.41(0.05) \par 6.51(0.06) \par 6.53(0.06) \par 6.54(0.06) \par
6.54(0.06)& 14.97(0.08) \par 10.28(0.07) \par 9.27(0.07) \par
8.52(0.07) \par 7.98(0.06) \par 6.93(0.06)& 6.00(1.82) \par
6.56(0.20) \par 6.56(0.10) \par 6.55(0.07) \par 6.55(0.06) \par
6.54(0.06)&
6.527 \\
\hline -0.5& 1 \par 8 \par 16 \par 32 \par 64 \par 1024&
2.57(0.02) \par 1.47(0.02) \par 1.42(0.02) \par 1.41(0.02) \par
1.39(0.02) \par 1.38(0.02)& 1.70(0.01) \par 1.41(0.02) \par
1.40(0.02) \par 1.40(0.02) \par 1.39(0.02) \par 1.38(0.02)&
0.67(0.01) \par 1.40(0.02) \par 1.40(0.02) \par 1.40(0.02) \par
1.39(0.02) \par 1.38(0.02)& 7.86(0.05) \par 3.63(0.04) \par
2.88(0.03) \par 2.45(0.03) \par 2.09(0.03) \par 1.55(0.03)&
1.62(0.96) \par 1.43(0.06) \par 1.41(0.03) \par 1.41(0.02) \par
1.39(0.02) \par 1.38(0.02)&
1.395 \\
\hline 1.0& 1 \par 8 \par 16 \par 32 \par 64 \par 1024&
11.36(0.06) \par 11.36(0.07) \par 11.37(0.07) \par 11.35(0.07)
\par 11.34(0.07) \par 11.33(0.07)& 8.05(0.05) \par 10.22(0.07)
\par 10.63(0.07) \par 10.84(0.07) \par 10.98(0.07) \par
11.24(0.07)& 6.31(0.05) \par 10.00(0.07) \par 10.49(0.07) \par
10.74(0.07) \par 10.91(0.07) \par 11.22(0.07)& 16.79(0.08) \par
14.35(0.08) \par 13.63(0.08) \par 13.06(0.07) \par 12.63(0.07)
\par 11.69(0.07)& 8.84(2.58) \par 10.68(0.74) \par 10.93(0.51)
\par 11.04(0.37) \par 11.12(0.29) \par 11.28(0.12)&
11.315 \\
\hline -1.0& 1 \par 8 \par 16 \par 32 \par 64 \par 1024&
0.415(0.002) \par 0.018(0.001) \par 0.014(0.001) \par 0.014(0.001)
\par 0.013(0.001) \par 0.013(0.001)& 0.167(0.001) \par
0.014(0.001) \par 0.013(0.001) \par 0.014(0.001) \par 0.013(0.001)
\par 0.013(0.001)& 0(0) \par 0.014(0.001) \par 0.013(0.001) \par
0.014(0.001) \par 0.013(0.001) \par 0.013(0.001)& 2.839(0.018)
\par 0.476(0.008) \par 0.250(0.06) \par 0.137(0.004) \par
0.080(0.003) \par 0.023(0.002)& 0.207(0.209) \par 0.016(0.003)
\par 0.014(0.001) \par 0.014(0.001) \par 0.013(0.001) \par
0.013(0.001)&
0.0131 \\
\hline
\end{tabular}

\caption{Two asset down-and-out call with lower barriers for the
first and second assets. $h_1 = h_2 = 90$, $S_1 (0) = S_2 (0) =
100$, $K = 100$, $\sigma _1 = \sigma _2 = 0.3$, $r = 0.1$, $T =
1$, $100\,000$ simulations. $Q_c $ is the exact value of the
option with continuously monitored barriers.}

\label{tab3}
\end{center}
\end{table}

\begin{table}[p]
\begin{center}
\begin{tabular}
{|p{16pt}|p{25pt}|p{60pt}|p{60pt}|p{60pt}|p{60pt}|p{60pt}|p{60pt}|}
\hline $d$ & $M$ & $Q_U $(std.err.)& $Q_I $(std.err.)& $Q_L
$(std.err.)& $Q_S $(std.err.)& $Q_2 $(std.err.)&
$Q_0 $(std.err.) \\
\hline 3 \par & 1 \par 2
\par 4
\par 8 \par 16 \par 32 \par 64
\par 1024& 8.96(0.07) \par 8.26(0.07) \par 7.83(0.07) \par
7.65(0.07) \par 7.60(0.08) \par 7.60(0.08) \par 7.60(0.08) \par
7.60(0.08)& 6.69(0.06) \par 7.20(0.07) \par 7.43(0.07) \par
7.51(0.07) \par 7.56(0.08) \par 7.59(0.08) \par 7.59(0.08) \par
7.60(0.08)& 5.13(0.06) \par 6.76(0.07) \par 7.31(0.07) \par
7.47(0.07) \par 7.54(0.08) \par 7.58(0.08) \par 7.59(0.08) \par
7.60(0.08)& 14.96(0.10) \par 13.27(0.09) \par 11.81(0.09) \par
10.76(0.09) \par 9.96(0.09) \par 9.29(0.08) \par 8.80(0.08) \par
7.91(0.08)& 7.83(1.20) \par 7.73(0.60) \par 7.63(0.27) \par
7.58(0.14) \par 7.58(0.10) \par 7.59(0.08) \par 7.59(0.08) \par
7.60(0.08)&
7.04(1.97) \par 7.51(0.82) \par 7.57(0.33) \par 7.56(0.16) \par 7.57(0.11) \par 7.59(0.09) \par 7.59(0.08) \par 7.60(0.08) \\
\hline 10 \par & 1 \par 2 \par 4 \par 8 \par 16 \par 32 \par 64
\par 1024& 4.62(0.05) \par 3.56(0.05) \par 2.98(0.05) \par
2.80(0.05) \par 2.71(0.05) \par 2.67(0.05) \par 2.65(0.05) \par
2.65(0.05)& 1.19(0.02) \par 1.97(0.03) \par 2.39(0.04) \par
2.60(0.05) \par 2.64(0.05) \par 2.65(0.05) \par 2.64(0.05) \par
2.65(0.05)& 0.21(0.01) \par 1.33(0.03) \par 2.20(0.04) \par
2.54(0.05) \par 2.61(0.05) \par 2.64(0.05) \par 2.64(0.05) \par
2.65(0.05)& 10.36(0.09) \par 7.92(0.08) \par 6.13(0.07) \par
5.09(0.07) \par 4.37(0.06) \par 3.84(0.06) \par 3.48(0.06) \par
2.86(0.05)& 2.90(1.75) \par 2.77(0.84) \par 2.68(0.34) \par
2.70(0.15) \par 2.67(0.08) \par 2.66(0.06) \par 2.65(0.05) \par
2.65(0.05)&
2.41(2.23) \par 2.45(1.16) \par 2.59(0.44) \par 2.67(0.18) \par 2.66(0.10) \par 2.66(0.07) \par 2.64(0.06) \par 2.65(0.05) \\
\hline
\end{tabular}

\caption{Knock-out calls on three and ten assets with lower
barriers for each of the asset. $S_i (0) = 100$, $h_i = 80$, $\rho
_{ij} = 0.5$ if $i \ne j$, $\sigma _i = 0.4$ ($i,j = 1,...,d)$, $T
= 1$, $K = 100$, $r = 0.05$, $100\,000$ simulations.}

\label{tab4}
\end{center}
\end{table}

\begin{figure}[p]
\centerline{\includegraphics[width=3.94in,height=1.97in]{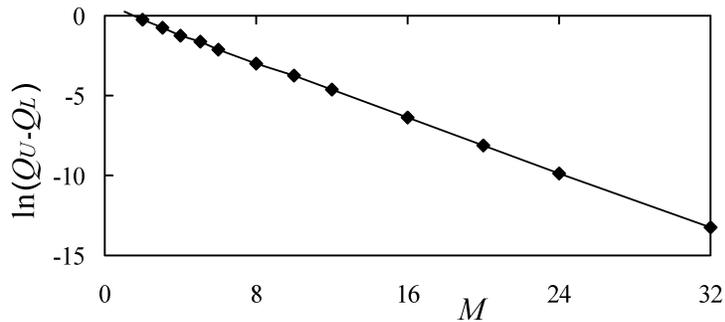}}
\caption{Double knock-out call on a single asset considered in
Table 2.}

\label{fig1}
\end{figure}

\begin{figure}[p]
\centerline{\includegraphics[width=3.94in,height=1.97in]{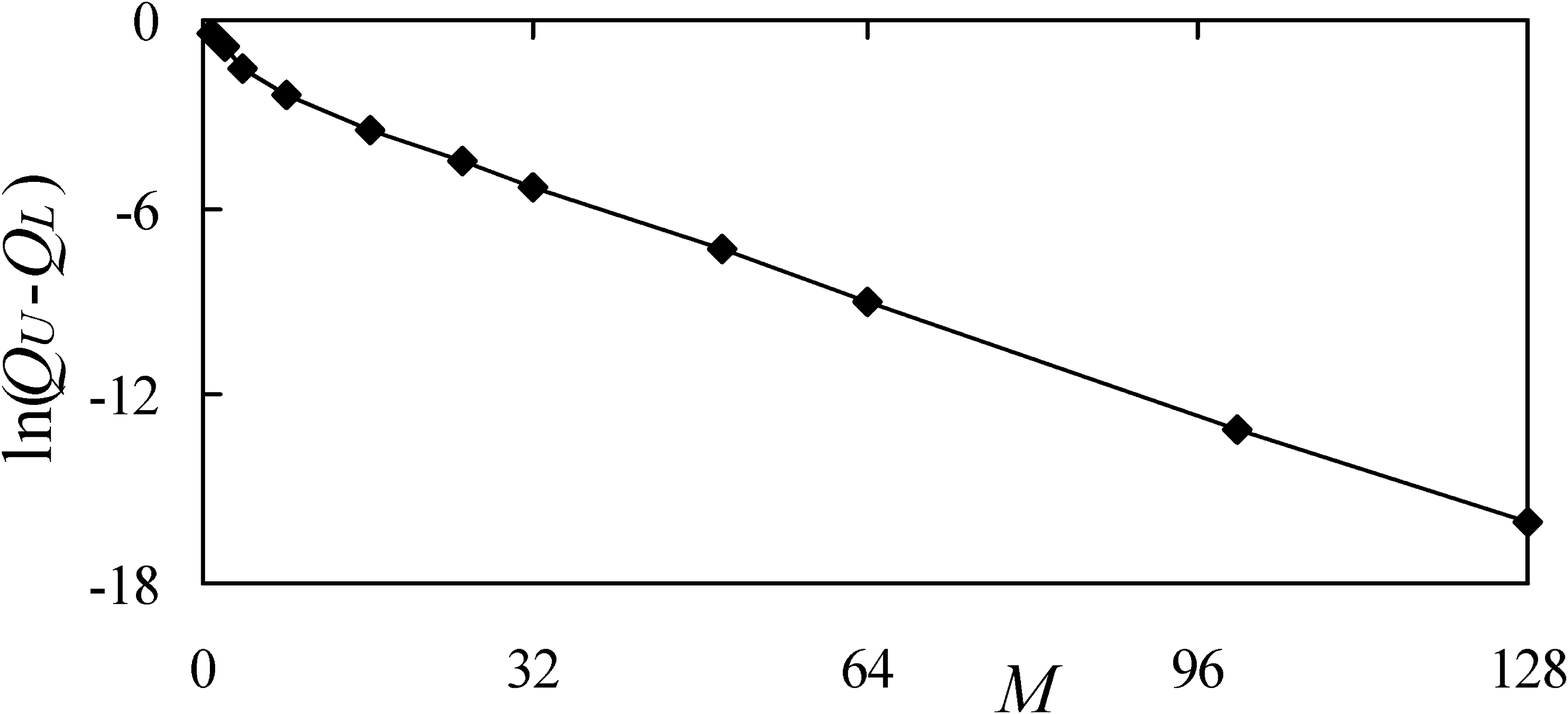}}

\caption{Two asset down-and-out call considered in Table 3 with
$\;\rho = - 1$.}

\label{fig2}
\end{figure}

\begin{figure}[p]
\centerline{\includegraphics[width=3.94in,height=1.97in]{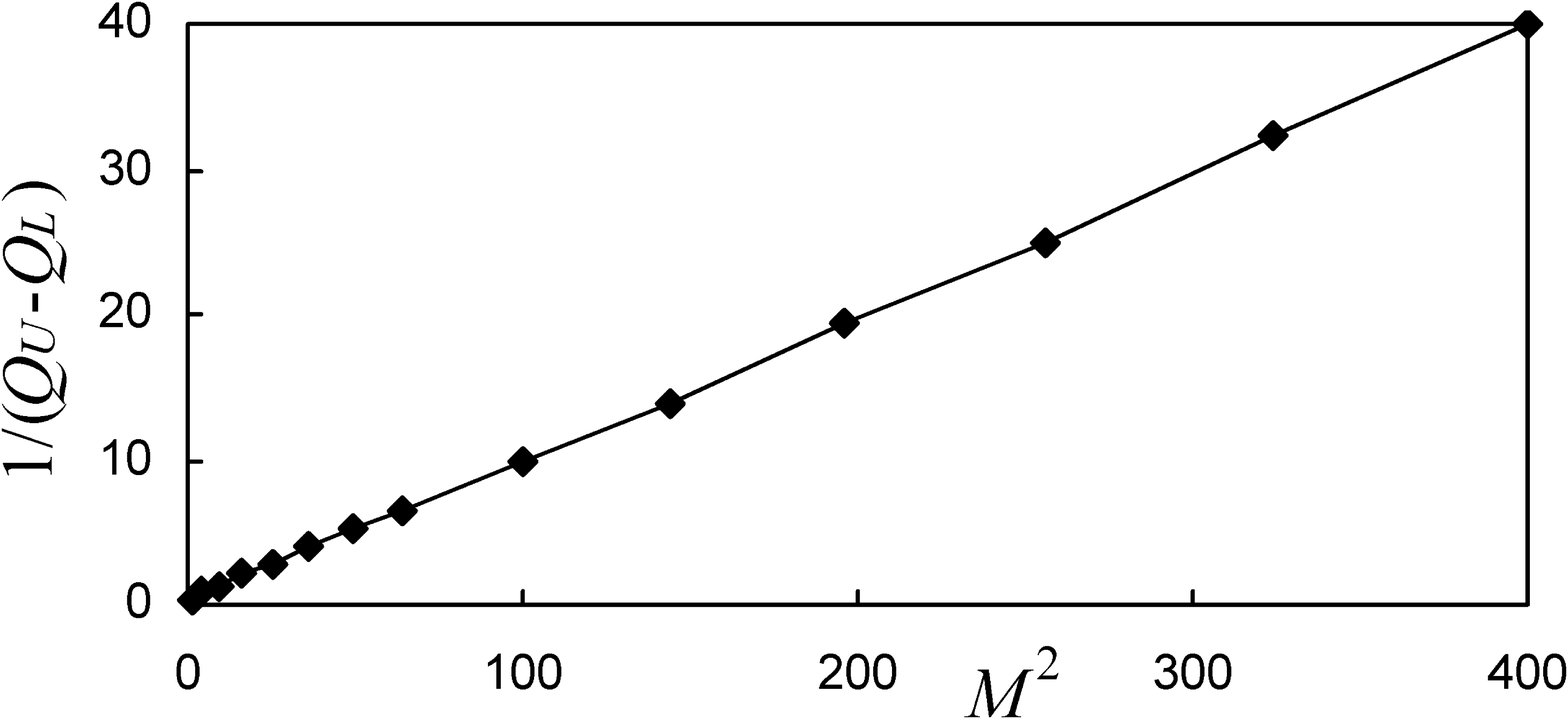}}

\caption{Two asset down-and-out call considered in Table 3 with
$\rho = 0$.}

\label{fig3}
\end{figure}

\begin{figure}[p]
\centerline{\includegraphics[width=3.94in,height=1.97in]{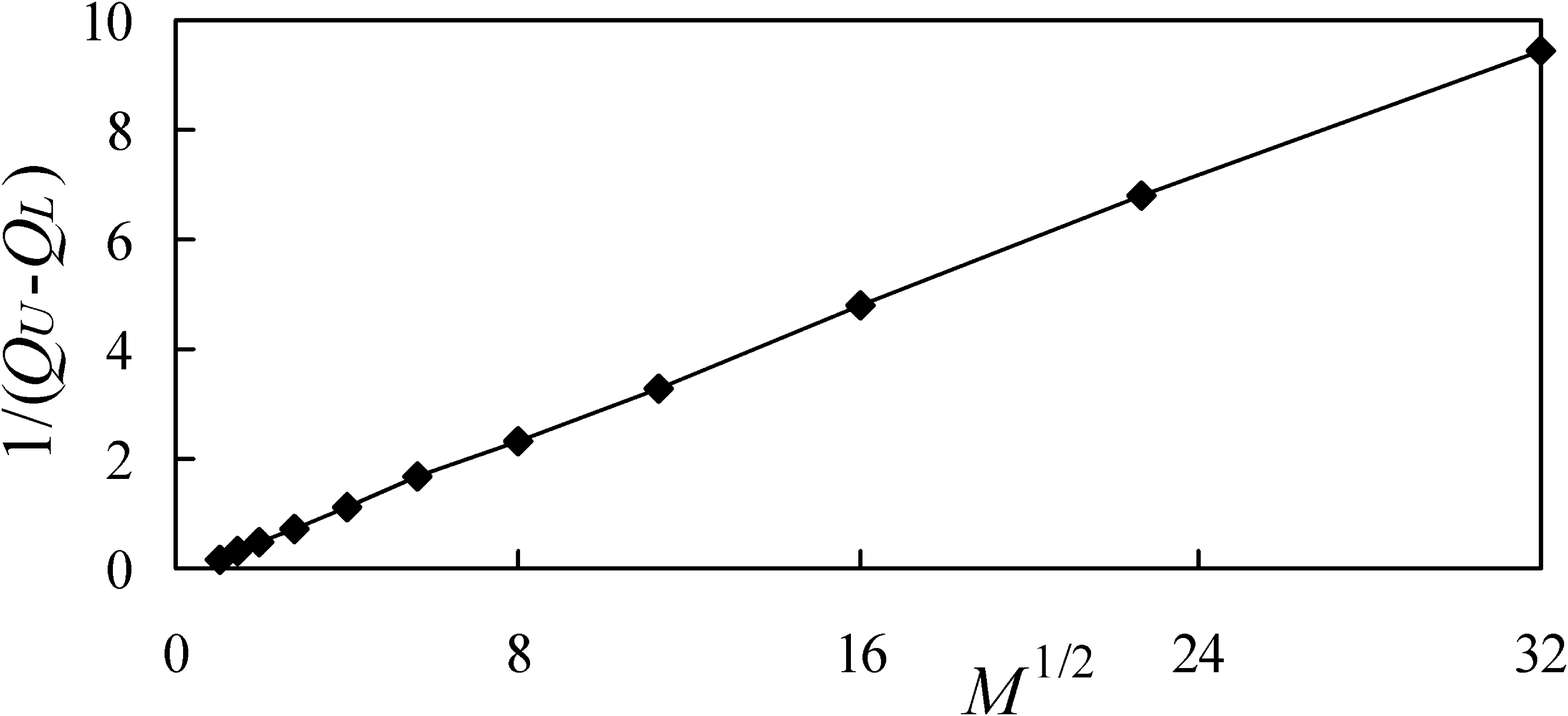}}

\caption{Two asset down-and-out call considered in Table 3
with$\;\rho = 1$.}

\label{fig4}
\end{figure}

\begin{figure}
\centerline{\includegraphics[width=3.94in,height=1.97in]{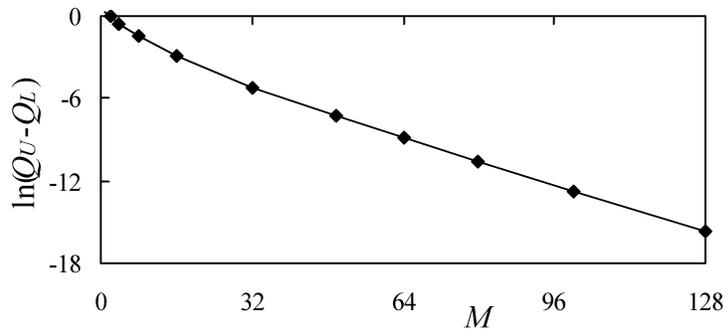}}

\caption{Two asset down-and-out call considered in Table 3 with
$\rho = 1$ and $S_1 (0) = 95$, $S_2 (0) = 105$.}

\label{fig5}
\end{figure}

\end{document}